\documentclass[twocolumn,showpacs,preprintnumbers,amsmath,amssymb]{revtex4}
\usepackage[dvips]{graphicx}
\usepackage{dcolumn}
\usepackage{bm}
\usepackage{color}
\usepackage{amsmath}

\def\cor#1{ #1 }     

\def\ket#1{ $ \left\vert  #1   \right\rangle $ }
\def\ketm#1{  \left\vert  #1   \right\rangle   }
\def\bram#1{  \left\langle  #1   \right\vert   }
\def\sprm#1#2{\left\langle #1 \left\vert \right. #2 \right\rangle}

\begin{document}

\title{High-fidelity copies from a symmetric $1\rightarrow2$
       quantum cloning machine}

\author{Michael Siomau}
\email{siomau@physi.uni-heidelberg.de}
\affiliation{Max-Planck-Institut f\"{u}r Kernphysik, Postfach 103980,
             D-69029 Heidelberg, Germany}
\affiliation{Physikalisches Institut, Heidelberg Universit\"{a}t,
             D-69120 Heidelberg, Germany}

\author{Stephan Fritzsche}
\affiliation{Department of Physical Sciences, P.O.~Box 3000,
             Fin-90014 University of Oulu, Finland}
\affiliation{Frankfurt Institute for Advanced Studies,
             D-60438 Frankfurt am Main, Germany}

\begin{abstract}
A symmetric $1\rightarrow2$ quantum cloning machine (QCM) is
presented that provides high-fidelity copies with $0.90\,\le\,
F\,\le\,0.95$ for all pure (single-qubit) input states from a given
meridian of the Bloch sphere. \cor{Emphasize is placed especially on
the states of the (so-called) Eastern meridian, that includes the
computational basis states $\ketm{0},\,\ketm{1}$ together with the
diagonal state $\ketm{+} \,=\, \frac{1}{\sqrt{2}} (\ketm{0} \,+\,
\ketm{1})$, for which suggested cloning transformation is shown to
be optimal.} In addition, we also show how this QCM can be utilized
for eavesdropping in Bennett's B92 protocol for quantum key
distribution with a substantial higher success rate than obtained
for universal or equatorial quantum copying.
\end{abstract}


\maketitle

\section{Introduction}

The ability to copy information is fundamental for many processes in
distributing and dealing with data. While classical information can
often be copied many times without any significant loss, the cloning
of quantum information is seriously restricted by the laws of
quantum mechanics. This limitation is known today as the
\textit{non-cloning} theorem \cite{Wootters:82} and has found its
applications in quite different fields of quantum information
theory, such as quantum computation \cite{Nielsen:00,Kilin:07} and
quantum cryptography \cite{Kilin:07,Scarani:05}.

In processing quantum information, formally, any device that
provides for $M$ \textit{unknown} input states of a given system $N
\;(>M)$ output states (of the same or some analogue systems) is
called a $M \rightarrow N$ quantum cloning machine (QCM). To ensure
a proper quantum behaviour, such machines are usually represented as
unitary transformations; they are called \textit{symmetric} if the
$N$ output states are identical to each other, and are said to be
\textit{nonsymmetric} otherwise. In this work, we shall consider
symmetric $1\rightarrow2$ QCM that provides for a (pure) input state
\ket{s_a} (of qubit $a$) the two qubits $a$ and $b$ in the same
output state $\rho_a^{out} \,=\,\rho_b^{out}$, and where
$\rho^{out}$ denotes the (reduced) density matrix of the
corresponding qubit. For an ideal transformation, of course, we
might expect
\begin{equation}
\label{int-1}
   \ketm{s}_a\ketm{0}_b\ketm{Q}_c \: \longrightarrow \:
   \ketm{s}_a\ketm{s}_b\ketm{\check{Q}}_c \, ,
\end{equation}
i.e.\ that the input and output states of qubit $a$ are the same,
$\ketm{s_a} \,=\, \ketm{s_a'}$, but this would be in conflict with
the non-cloning theorem mentioned above. In the transformation
(\ref{int-1}), as usual, we here suppose that the qubit $b$ was
initially prepared in the state $\ketm{0}_b$ and that the state
vectors $\ketm{Q}_c$ and $\ketm{\check{Q}}_c$ denote the initial and
final states of the corresponding cloning device. In order to
characterize the `quality' of a cloning device, we shall often use
below the fidelity
\begin{equation}
   F \,=\,  \sprm{s|\rho^{out}}{s} \, ,
\label{int-3}
\end{equation}
between the input and output state (of qubit $a$), which takes the
values $0 \,\le\, F\,\le\, 1$ and where $F\,=\,1$ refers to the case
that $\rho^{out} \,\equiv\, \ketm{s}\bram{s}$.

Several QCM's have been discussed in the literature before
\cite{Scarani:05,Buzek:96,Cerf:98,Bruss:00}; Bu\v{z}ek and Hillery
\cite{Buzek:96}, for example, worked out a symmetric $1\rightarrow2$
QCM that provides copies with the fidelity $F=5/6 \approx 0.83$
independent of the given input state. \cor{This state-independent
transformation, which is now known as \textit{universal} QCM, has
found recent attention and application in quantum cryptography,
namely, in optimal eavesdropping attack \cite{Bechmann:99} within
the six-state protocol \cite{Bruss:98}.} For some protocols in
quantum cryptography, however, only few states $\{\ketm{s_a}\}$ are
(pre-) selected from the Bloch sphere and, hence, one may wish to
find a state-dependent QCM that provides copies with higher fidelity
than the universal Bu\v{z}ek-Hillery machine for this particular set
of states.

The best-known example of such a state-dependent QCM is
\textit{equatorial} (or phase-covariant) QCM \cite{Bruss:00}. This
QCM provides copies with fidelity $F = 1/2 + \sqrt{1/8} \,\approx\,
0.85$ for all states which are taken from the equatorial plane of
the Bloch sphere. Equatorial QCM has also found a remarkable
application in quantum cryptography, since it can be used to improve
the efficiency of an eavesdropping attack within the BB84 protocol
\cite{Bennett:84}. While, in fact, only four states (from the
equatorial plane) are used in order to encode the information in the
BB84 protocol, equatorial QCM enables an eavesdropper to obtain
copies of these states with the highest fidelity, if compared with
any other known QCM \cite{Scarani:05}.

In this work, we here present and discuss another example of
state-dependent QCM --- a (meridional) QCM  that provides copies
with high-fidelity $0.9 \,\le\, F \,\le\, 0.95$ for all states along
a given half-circle (the `Eastern' meridian) of the Bloch sphere,
\cor{that includes three states $\{ \ketm{0},\: \ketm{1}, \:
\ketm{+} \}$ which define this meridian uniquely. This newly
suggested QCM is constructed to be optimal for copying of all input
states from the meridian.} We also analyze how this newly suggested
QCM can be applied in quantum cryptography for the eavesdropping
within Bennett's B92 quantum key distribution protocol
\cite{Bennett:92}. It is shown, in particular, that this
transformation provides a substantial higher success rate than
obtained for a universal or equatorial quantum copying.

The paper is organized as follows. In the next section, let us first
introduce and discuss the general form of symmetric $1\rightarrow2$
cloning transformations. In particular, here we shall analyze the
Bu\v{z}ek-Hillery-type transformations \cite{Buzek:96} which can be
utilized for both, a state-dependent and state-independent
(symmetric) copying of quantum states. By making use of some
additional parameter in the definition of the cloning transformation
(when compared to Ref.~\cite{Buzek:96}), we then present a QCM in
Subsection~\ref{subsec:2.b} which enables one to produce \cor{for a
selected set of pure input states from the Bloch sphere copies with
higher fidelity than it can be obtained for universal or equatorial
quantum copying. The QCM is than defined, in
Subsection~\ref{subsec:2.c}, to be optimal for copying states from
chosen meridian of the Bloch sphere.} Below, therefore, we will
refer to this cloning transformation as a \textit{meridional} QCM,
even if the particular region of high-fidelity cloning can be chosen
rather freely by adapting just three parameters in the given cloning
transformation. \cor{In Subsection~\ref{subsec:2.d} we present
explicit form of the meridional cloning transformation and recall
properties of it.} In Section~\ref{sec:3}, we later analyze for this
new QCM the success (or failure) of an potential eavesdropper within
Bennett's B92 protocol \cite{Bennett:92} for the distribution of
quantum keys. In this section, emphasis is placed especially on the
question how efficient such an eavesdropper can attack the
transmission of quantum information; in our discussion below, this
will be quantified by means of the mutual information between the
eavesdropper and the legitim user as well as the discrepancy of a
qubit (if compared with the originally transmitted one) after a
corresponding attack has been made. Finally, a few conclusions are
drawn in Section~\ref{sec:4}.

\section{\label{sec:2} Quantum cloning for a restricted set of input states}

\subsection{\label{subsec:2.a} General symmetric $1\rightarrow2$ cloning
            transformations}

To introduce some notations that are necessary for our discussion
below, let us start from a single qubit $a$ whose (pure) states can
be written in the Bloch sphere representation as $\ketm{s}_a \,=\,
\cos{\frac{\theta}{2}} \ketm{0}_a +
                  \sin{\frac{\theta}{2}} e^{i\phi}\ketm{1}_a$,
and with $\ketm{0}_a$ and $\ketm{1}_a$ being the standard
(computational) basis states. In this representation, the parameters
$\theta$ and $\phi$ take their values from $0 \,\le\, \theta \,\le\,
\pi$ and $0 \,\le\, \phi \,<\, 2\pi$, respectively, and we shall
often use this Bloch sphere (picture) in order to visualize the
states of interest. Moreover, let us refer to the intersection of
the Bloch sphere with the \textit{x-z} plane as the \textit{main
circle}, so that all states from this intersection can be
parameterized by means of just the (single) parameter $\theta$ as
\begin{eqnarray}
\label{main-meridian}
   \ketm{s}_a & = & \cos{\frac{\theta}{2}} \ketm{0}_a
                \pm \sin{\frac{\theta}{2}}\ketm{1}_a \, .
\end{eqnarray}
While, in this expression, the `+' sign refers to the right
(Eastern) meridian of the main circle and includes for $\theta =
\pi/2$ also the diagonal state $\ketm{+} \;=\; \frac{1}{\sqrt{2}}
\,\left( \ketm{0} + \ketm{1} \right)$, the `--' sign is associated
with its left (Western) meridian and includes the state $\ketm{-}$.
For the sake of simplicity in visualizing the different states on
the Bloch sphere, we shall sometimes use also this `geographical'
notation in our discussion below.

Although the state $\ketm{s}_a$ of a given qubit will be typically
in a superposition of the two basis states $\ketm{0}_a$ and
$\ketm{1}_a$, it is of course sufficient to know the transformation
of just the basis in order to obtain a proper copying of states.
Therefore, the most general $(1 \rightarrow 2)$ quantum cloning
transformation for the state of qubit $a$ upon qubit $b$ can be cast
into the form
\begin{eqnarray}
\label{general-QCT}
   \ketm{0}_a \ketm{0}_b \ketm{Q}_c  &\longrightarrow &
   \sum_{m,n=0}^1 \, \ketm{m}_a \ketm{n}_b \ketm{Q_{mn}}_c \, ,
\nonumber \\[0.1cm]
   \ketm{1}_a \ketm{0}_b \ketm{Q}_c  &\longrightarrow &
   \sum_{m,n=0}^1 \, \ketm{m}_a \ketm{n}_b \ketm{\tilde{Q}_{mn}}_c \, ,
\end{eqnarray}
where $\ketm{Q}_c$ denotes the initial state of the cloning
apparatus, and where we have assumed --- without loss of generality
--- that the second qubit $b$ was prepared initially in the basis
state $\ketm{0}_b$. Once, the transformation has been performed,
$\ketm{m}_a$ and $\ketm{n}_b$ denote the output basis states of the
two copies, while $\ketm{Q_{mn}}_c$ and $\ketm{\tilde{Q}_{mn}}_c$
are the corresponding states of the apparatus. As seen from the
transformation (\ref{general-QCT}), here we do not assume any
additional condition for the final states of the cloning apparatus.
However, in order to ensure that the transformation
(\ref{general-QCT}) is \textit{unitary},
\begin{eqnarray}
\label{unitary-cloning}
   \sum_i \, c_i \ketm{i}_{ab} \ketm{Q}_c &\longrightarrow&
   \sum_{i,\lambda} \, c_i U_{i\lambda}\ketm{\lambda}_{abc} \, ,
\end{eqnarray}
for all possible input states, i.e.\ for $\ketm{i}_{ab} \,=\,
\{\ketm{0}_a\ketm{0}_b,$ $ \ketm{1}_a\ketm{0}_b \}$, the final
states of the apparatus must fulfill certain requirements
\cite{Peres:00}.

In Eq.~(\ref{unitary-cloning}), the three-partite basis $\{
\ketm{\lambda}_{abc} \}$ refers to a complete and orthonormal basis
for the overall system `qubits a,b + apparatus'. Thus, the requested
unitarity $UU^\dag \,=\, 1$ of the transformation
(\ref{general-QCT}) implies the conditions
\begin{eqnarray}
\label{general-cond}
   \sum_{m,n=0}^1  \, \sprm{Q_{mn}}{Q_{mn}} & = &
   \sum_{m,n=0}^1  \, \sprm{\tilde{Q}_{mn}}{\tilde{Q}_{mn}} \; = \;
   1 \, ,
   \nonumber  \\
   \sum_{m,n=0}^1  \, \sprm{Q_{mn}}{\tilde{Q}_{mn}} & = & 0 \, .
\end{eqnarray}
For any explicit construction of a $(1 \rightarrow 2)$ quantum
cloning transformation, we must therefore `determine' the final
states $\ketm{Q_{mn}}_c$ and $\ketm{\tilde{Q}_{mn}}_c$ of the
apparatus in line with the conditions (\ref{general-cond}). These
state vectors then define the QCM uniquely.

Before we shall construct QCM with some particular properties, let
us consider the simplest case of such a transformation
(\ref{general-QCT}) as first suggested by Wootters and Zurek
\cite{Wootters:82}
\begin{eqnarray}
\label{Wootters/Zurek-QCT}
   \ketm{0}_a \ketm{0}_b \ketm{Q}_c  &\longrightarrow &
   \ketm{0}_a \ketm{0}_b \ketm{Q_{00}}_c \, ,
\nonumber\\[0.2cm]
   \ketm{1}_a \ketm{0}_b \ketm{Q}_c  &\longrightarrow &
   \ketm{1}_a \ketm{1}_b \ketm{\tilde{Q}_{11}}_c \, .
\end{eqnarray}
In this transformation, obviously, only a single term is retained
from the summations on the right-hand side (rhs) of
Eqs.~(\ref{general-QCT}), i.e.\ the term with $m = n = 0$ in the
first and $m = n = 1$ in the second line. No further freedom remains
in the set-up of this transformation since the unitarity conditions
(\ref{general-cond}) then require $\sprm{Q_{00}}{Q_{00}} =
\sprm{Q_{11}}{Q_{11}} = 1$. As seen from
Eq.~(\ref{Wootters/Zurek-QCT}), moreover, the Wootters-Zurek
transformation is symmetric with regard to an interchange of the
basis states $\ketm{0}_a \leftrightarrow \ketm{1}_a$ and this
implies that the same symmetry helds also for the state of the two
copies $\rho_a^{out}$ and $\rho_b^{out}$, i.e. $\rho_a^{out}
\,=\,\rho_b^{out}$. For the Wootters-Zurek transformation,
furthermore, the fidelity between the input state (from the main
cirlce) and the corresponding output is given by
\begin{eqnarray}
\label{fidelity W-Z QCT}
   F(\theta) & = & 1 - \frac{1}{2}\sin^2{\theta} \, .
\end{eqnarray}
Therefore, this particular transformation can provide an `exact'
copy with fidelity $F(0) \,=\, F(\pi) \,=\,1$ just for the basis
states \ket{0} and \ket{1}, i.e.\ the two `poles' of the Bloch
sphere, while the fidelity drops down to $F(\frac{\pi}{2}) \,=\,
1/2$ for all states along the equator and, especially, for the two
diagonal states \ket{\pm}. For this reason, the Wootters-Zurek
transformation appears to be of little help if one wishes to copy
(unknown) states other than the basis states themselves.

\subsection{\label{subsec:2.b} Cloning transformations of
            Bu\v{z}ek-Hillery type}

Additional terms in the cloning transformation (\ref{general-QCT})
need to be considered if we wish to construct a QCM that supports an
equal-fidelity cloning for all states on the Bloch sphere, or which
improves the fidelity between the input and output states for
certain regions on this sphere. For example, Bu\v{z}ek and Hillery
\cite{Buzek:96} have analyzed in quite detail the three-term
transformation
\begin{eqnarray}
\label{Buzek/Hillery-QCT-1}
   \ketm{0}_a \ketm{0}_b \ketm{Q}_c
   &\longrightarrow &  \ketm{0}_a \ketm{0}_b \ketm{Q_0}_c
\nonumber\\[0.1cm]
   & & \hspace*{-0.05cm} \:+\:
   \left[ \ketm{0}_a \ketm{1}_b  + \ketm{1}_a \ketm{0}_b
   \right] \, \ketm{Y_0}_c \, ,
\\[0.2cm]
   \ketm{1}_a \ketm{0}_b \ketm{Q}_c
   &\longrightarrow &  \ketm{1}_a \ketm{1}_b \ketm{Q_1}_c
\nonumber\\[0.1cm]
\label{Buzek/Hillery-QCT-2}
   & & \hspace*{-0.05cm} \:+\:
   \left[ \ketm{0}_a \ketm{1}_b + \ketm{1}_a \ketm{0}_b
   \right] \, \ketm{Y_1}_c , \quad
\end{eqnarray}
which takes into account two additional terms (compared with the
Wootters-Zurek transformation (\ref{Wootters/Zurek-QCT})) and which
gives us further freedom in choosing the final states of the cloning
apparatus. To follow the notation by Bu\v{z}ek and Hillery
\cite{Buzek:96}, here we have introduced $\ketm{Q_0} \equiv
\ketm{Q_{00}}$, $\,\ketm{Y_0} \equiv \ketm{Q_{01}} = \ketm{Q_{10}}$,
$\,\ketm{Q_1}
              \equiv \ketm{\tilde{Q}_{11}}$,  and
$\,\ketm{Y_1} \equiv \ketm{\tilde{Q}_{01}} = \ketm{\tilde{Q}_{10}}$
to denote the final states of the apparatus. Again, in order to
ensure the symmetry with regard to an interchange $\ketm{0}_a
\leftrightarrow \ketm{1}_a$ of the basis states of qubit $a$, we
have assumed conditions $\ketm{Q_{01}} = \ketm{Q_{10}}$ and
$\ketm{\tilde{Q}_{01}} = \ketm{\tilde{Q}_{10}}$ for the final states
of the cloning apparatus.

Eqs.~(\ref{Buzek/Hillery-QCT-1})-(\ref{Buzek/Hillery-QCT-2}) can be
utilized to define cloning transformations for which the fidelity
between the input and output states is either state-independent or
depends explicity on the given input. This behaviour depends on the
additional restriction we shall place on the final states of the
apparatus, beside of the unitarity conditions
(\ref{unitary-cloning}). For the sake of simplicity, let us omit in
the following the indices $a,\, b$ and $c$ of the individual
subsystems but keep in mind that $\ketm{kl} \,\equiv\, \ketm{k}_a
\ketm{l}_b$ always refers to the state of the two copies to be
created, and that the vectors \ket{Q_i} and \ket{Y_i} belong to the
cloning apparatus. With this change in the notation, we find from
Eqs.~(\ref{general-cond}) that the final-state vectors of the
apparatus in the Bu\v{z}ek-Hillery transformation must satisfy the
conditions
\begin{eqnarray}
\label{unit-con-12}
   \sprm{Q_i}{Q_i} + 2 \sprm{Y_i}{Y_i} & = & 1 \, ,
   \quad {\rm for} \quad i \:=\: 0,1 \, ;
   \\[0.1cm]
\label{unit-con-3}
   \sprm{Y_0}{Y_1} & = & 0 \, ,
\end{eqnarray}
in order to ensure the unitarity of the transformation. For a
general (pure) input state (\ref{main-meridian}) from the main
circle, the QCM
(\ref{Buzek/Hillery-QCT-1})-(\ref{Buzek/Hillery-QCT-2}) then gives
rise to the two-qubit density operator $\rho_{ab}^{\rm out}$ which
contain 14 scalar products between the final-state vectors from the
apparatus. Each scalar product introduces a (complex) parameter for
the Bu\v{z}ek-Hillery transformation
(\ref{Buzek/Hillery-QCT-1})-(\ref{Buzek/Hillery-QCT-2}). This gives
rise to a total of 14 parameters for the quantum cloning
transformation
(\ref{Buzek/Hillery-QCT-1})-(\ref{Buzek/Hillery-QCT-2}), while only
the 3 restrictions (\ref{unit-con-12})-(\ref{unit-con-3}) need to be
fulfilled due to the unitarity of the transformation.

Further conditions must therefore be formulated in order to define
the QCM properly. For example, we may choose all scalar products to
be real and also request that the two final-state vectors \ket{Y_0}
and \ket{Y_1} have an equal norm
\begin{eqnarray}
\label{notations-1}
   \sprm{Y_0}{Y_0} & = & \sprm{Y_1}{Y_1} \;\equiv\; \zeta \, .
\end{eqnarray}
Under these conditions, Eq.~(\ref{unit-con-12}) takes the same form
for $i=0$ and $1$, and hence, $\sprm{Q_0}{Q_0} = \sprm{Q_1}{Q_1} = 1
- 2\zeta $. There are two other conditions that can be obtained from
Eqs.~(\ref{unit-con-12})-(\ref{unit-con-3}) by using the restriction
(\ref{notations-1}), namely,
\begin{eqnarray}
\label{notations-2}
   \sprm{Y_0}{Q_1} &=& \sprm{Y_1}{Q_0} \;\equiv\; \eta/2 \, ,
   \\[0.01cm]
\label{notations-3}
   \sprm{Q_1}{Y_1} &=& \sprm{Q_0}{Y_0} \;\equiv\; \kappa/2 \, .
\end{eqnarray}
With these additional notations and conditions, we have arrived at
the final-state density matrix $\rho_{ab}^{\rm out}$ of the
transformation
(\ref{Buzek/Hillery-QCT-1})-(\ref{Buzek/Hillery-QCT-2}) that now
depends only on three real parameters. This is quite in contrast to
the original work of Bu\v{z}ek and Hillery \cite{Buzek:96} who
assumed the final states of the apparatus $ \sprm{Q_i}{Y_i} \,=\, 0$
to be pairwise orthogonal to each other for the cloning of the two
basis states $\ketm{i}, \: i = 0,1$. While the condition
(\ref{notations-3}) introduces a nonorthogonality between the final
states of the apparatus, three parameters $\zeta, \eta$ and $\kappa$
enables us to provide high-fidelity copies for a region of input
states from the Bloch sphere. However, the three parameters
$\zeta,\,\eta$ and $\kappa$ are not completely independent of each
other but must fulfill the three inequalities
\begin{eqnarray}
\label{Schwarz-con-1}
    0 &\leq& \zeta  \:\leq\: \frac{1}{2} \, ,
\label{Schwarz-con-2}\\[0.01cm]
    0 &\leq& \eta   \:\leq\: 2\sqrt{\zeta(1-2\zeta)} \, ,
\label{Schwarz-con-3}\\[0.01cm]
    0 &\leq& \kappa \:\leq\: 2\sqrt{\zeta(1-2\zeta)} \, .
\end{eqnarray}
due to Schwarz' inequality for the state vectors of the cloning
apparatus.

Using the final-state density matrix $\rho_{ab}^{\rm out}$ for the
transformation
(\ref{Buzek/Hillery-QCT-1})-(\ref{Buzek/Hillery-QCT-2}), it is
simple to show that the reduced density operator
\begin{eqnarray}
\label{reduced-den-matrix}
   \rho^{\rm\, out} & = &  \left(\cos^2{\frac{\theta}{2}} -
   \zeta \cos{\theta} \right)  \ketm{0}\bram{0}
   \nonumber \\
   &  & +\;  \frac{1}{2}
   \left( \kappa \pm \eta \sin{\theta} \right) \:
          \left( \ketm{0}\bram{1} + \ketm{1}\bram{0} \right)
   \nonumber \\
   &  & +\;  \left( \sin^2{\frac{\theta}{2}} + \zeta \cos{\theta} \right)
   \ketm{1}\bram{1} \, ,
\end{eqnarray}
is the same for both subsystems $a$ and $b$, that is $\rho^{\rm out}
\:\equiv\: {\rm Tr}_a \,(\rho_{ab})
                     \:=\: {\rm Tr}_b \,(\rho_{ab})$.
We can utilize this expression (\ref{reduced-den-matrix}) to
calculate also the fidelity between the input and output for all
states $\ketm{s}_a \,=\, \cos{\frac{\theta}{2}} \ketm{0}_a \pm
\sin{\frac{\theta}{2}}\ketm{1}_a$ along the main circle
\begin{eqnarray}
\label{fidelity}
   F(\theta) & \equiv & \sprm{s|\rho^{out}}{s}
   \nonumber \\
   & = & (1 - \zeta) \, - \, \frac{1}{2} (1 - \eta - 2\zeta)
         \sin^2{\theta} \pm \frac{\kappa}{2} \sin{\theta} \, .
\end{eqnarray}
Although the parameters $\zeta,\,\eta$ and $\kappa$ are restricted
by the inequalities (\ref{Schwarz-con-1})-(\ref{Schwarz-con-3}), it
is this \textit{freedom} in choosing these parameters that enables
us to optimize the fidelity $F(\theta)$ for certain (regions of)
states. Since $\kappa$ has a different sign for the two parts of the
main circle, a nonzero value this parameter leads to a quite
different behavior of the fidelity along the Western and Eastern
meridian:the high fidelity along the Eastern meridian correspond to
positive parameter $\kappa$, while the high fidelity along the
Western meridian is achieved for negative $\kappa$. So, for proper
values of $\zeta, \eta$ and $\kappa$ we can obtain a high fidelity
for one meridian.

\subsection{\label{subsec:2.c}
           \cor{Optimization of the cloning transformation}   }

\cor{Eq.~(\ref{fidelity}) can be applied to determine an `optimal'
set of parameters, either for a few selected input states or for a
whole region of states from the Bloch sphere. General method for
optimization of parameters of the unitary transformation was
developed in Ref.~\cite{Audenaert:02} and was successfully applied
to show optimality of some cloning transformations
\cite{Fiurasek:03}. The optimal cloning transformation maximizes
average single-clone fidelity}
\begin{equation}
 \label{average fidelity}
\overline{F} = \int_\Omega \: \frac{d \theta}{N} \; F(\theta)
\end{equation}
\cor{for chosen region of states $\Omega$ on the Bloch sphere, where
$N$ is the normalization factor. Substituting in this expression the
fidelity function (\ref{fidelity}) and integrating over all states
from Eastern meridian of the Bloch sphere, we obtain average
fidelity as function of the parameters $\zeta, \eta$ and $\kappa$,
i.e.}
\begin{eqnarray}
 \label{explicit_fid}
 \overline{F} =\int_0^\pi \frac{d \theta}{\pi} F(\theta) =
 \frac{1}{4} \left( 3 - 2 \zeta + \eta + \frac{4 \kappa}{\pi}
 \right) \, .
\end{eqnarray}
\cor{Following numerical optimization procedure over the three
parameters, which are restricted with inequalities
(\ref{Schwarz-con-1})-(\ref{Schwarz-con-3}), gives values
\begin{eqnarray}
\label{fidelity-3}
   \kappa \:=\: \frac{2}{5},  \qquad
   \zeta  \:=\: \frac{1}{10}, \qquad
   \eta   \:=\: \frac{2}{5} \, ,
\end{eqnarray}
approximately, that correspond to the maximal average fidelity
$\overline{F}\approx 0.927$.}

\cor{There is, however, another way to perform optimization of the
fidelity function (\ref{fidelity}) that does not require numerical
calculations. It is known that parameters of universal QCM can be
found from the requirement of optimal copying of just the discrete
set of six states that lie on $x,y$ and $z$ axis of the Bloch sphere
\cite{Scarani:05}. Similarly the requirement of optimal copying of
four states that lie on $x$ and $y$ axis of the sphere is sufficient
to determine equatorial QCM. We may, for example, request an equal
and maximum fidelity for just three selected states from the
meridian in order to determine optimal parameters of the cloning
transformation
(\ref{Buzek/Hillery-QCT-1})-(\ref{Buzek/Hillery-QCT-2}). If the
input $\ketm{s}$ is taken from the set of the three states $\{
\ketm{0}, \ketm{1}, \ketm{+} \}$ and we request an equal-fidelity
cloning of them, the maximum fidelity $F = 0.90$ is obtained for the
parameters (\ref{fidelity-3}). This result is not surprising since
the fidelity has local minima for the states $\ketm{0},\, \ketm{1}$
and $\ketm{+}$; that is the main reason of optimization with this
three states. In fact, in this optimization procedure we restricted
the fidelity function (\ref{fidelity}) downwards.}

\subsection{\label{subsec:2.d} \cor{Explicit form and properties
                               of the cloning transformation}}

\cor{With the help of the parameters (\ref{fidelity-3}) the
final-state vectors of the apparatus can be defined as
\begin{eqnarray}
   \ketm{Y_0} &=& \left\{ \frac{1}{\sqrt{10}}, 0 \right\} \, ,\qquad
   \ketm{Y_1} =   \left\{ 0,\frac{1}{\sqrt{10}}\right\} \, ,
\nonumber \\[0.1cm]
\label{vectors-QCM}
   \ketm{Q_0} &=& \left\{ \sqrt{\frac{2}{5}}, \sqrt{\frac{2}{5}} \right\}
   \, , \quad
   \ketm{Q_1} =   \left\{ \sqrt{\frac{2}{5}}, \sqrt{\frac{2}{5}}
   \right\} \, ,
\end{eqnarray}
in line with the conditions (\ref{unit-con-12})-(\ref{unit-con-3})
and (\ref{notations-1})-(\ref{notations-3}) from above. These four
vectors span only a two-dimensional subspace within the
(four-dimensional) space of the general copying machine
(\ref{general-QCT}). If we introduce the orthogonal basis $\ketm{0}$
and $\ketm{1}$ for this subspace, the transformation
(\ref{Buzek/Hillery-QCT-1})-(\ref{Buzek/Hillery-QCT-2}) can be
brought into the form}
\begin{eqnarray}
   \ketm{0} \ketm{0} \ketm{Q} &\longrightarrow&
   \sqrt{\frac{2}{5}} \ketm{00}
         \left( \ketm{0} + \ketm{1} \right)
   \nonumber
   \\[0.05cm]
\label{suggested-QCM-1}
   &  & \hspace*{0.3cm}
   \;+\; \frac{1}{\sqrt{10}}
         \left( \ketm{01} + \ketm{10} \right)\, \ketm{0} \, ,
   \\[0.1cm]
   \ketm{1} \ketm{0} \ketm{Q}
   &\longrightarrow&
   \sqrt{\frac{2}{5}} \ketm{11}
         \left( \ketm{0} + \ketm{1} \right)
   \nonumber
   \\[0.05cm]
\label{suggested-QCM-2}
   &  & \hspace*{0.3cm}
   \;+\; \frac{1}{\sqrt{10}}
         \left( \ketm{01} + \ketm{10} \right)\, \ketm{1} \, ,
\end{eqnarray}
\cor{if the vectors (\ref{vectors-QCM}) are substituted into the
transformation
(\ref{Buzek/Hillery-QCT-1})-(\ref{Buzek/Hillery-QCT-2}). This makes
our suggested QCM now explicit. The QCM
(\ref{suggested-QCM-1})-(\ref{suggested-QCM-2}) is optimal for a
symmetric hight-fidelity cloning of the states from Eastern
meridian.}

\cor{If we substitute the parameters (\ref{fidelity-3}) into this
formula, the input-output fidelity for the states from the main
circle becomes
\begin{eqnarray}
   F(\theta)  = \; \frac{9}{10} - \frac{1}{5}\sin^2{\theta} \pm
   \frac{1}{5} \sin{\theta} \, .
\label{fidelity-suggested-QCM}
\end{eqnarray}
Figure~\ref{fig-1} displays the behavior of this fidelity for the
two meridians of the main circle. While the fidelity is $F\,\geq\,
0.9$ for all states along the Eastern meridian (associated with the
$+$ sign in Eq.~(\ref{fidelity-suggested-QCM})), it drops quickly
down up to $F\,=\, 1/2$ for the \ket{-} state along the Western
meridian. Along the Eastern part, that includes the three states $\{
\ketm{0}, \ketm{1}, \ketm{+} \}$ from above, however, a remarkably
small variation of the fidelity with $0.9 \,\le\, F \,\le\, 0.95$
occurs and may favor this region to produce quantum copies with high
fidelity.}
\cor{
\begin{figure}
\begin{center}
\includegraphics[scale=0.8]{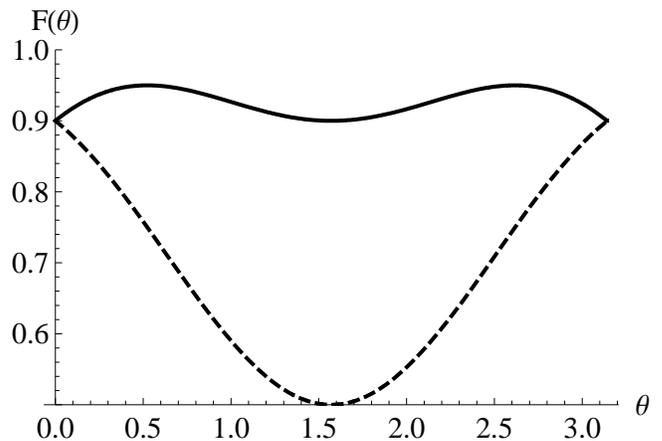}
\caption{Fidelity (\ref{fidelity-suggested-QCM}) of the quantum
cloning transformation
(\ref{suggested-QCM-1})-(\ref{suggested-QCM-2}) as function of the
angle $\theta$. The fidelity between the input and output states are
shown for the states (\ref{main-meridian}) from the main circle;
Eastern meridian (solid line) and Western meridian (dashed line).
See text for further discussion.} \label{fig-1}
\end{center}
\end{figure}}

\cor{So far, we have considered input states from the main circle
for the transformation
(\ref{suggested-QCM-1})-(\ref{suggested-QCM-2}). If we apply this
transformation to other (pure) states from the Bloch sphere
$\ketm{s}_a \,=\, \cos{\frac{\theta}{2}} \ketm{0}_a
                + \sin{\frac{\theta}{2}} e^{i\phi}\ketm{1}_a$ with
$\phi \neq 0$, the fidelity
\begin{eqnarray}
   F(\theta, \phi) =\; \frac{9}{10} - \frac{1}{5} \sin{\theta} \left(
   \sin{\theta} - \cos{\phi} \right) \, ,
\label{fidelity-final}
\end{eqnarray}
becomes dependent on the angle $\phi$ also and makes the behavior
slightly more complex. Note that this particular dependence
(\ref{fidelity-final}) of the fidelity was obtained from the request
that the states from Eastern meridian can be copied in optimal way.
By making similar requirements for other meridians on the Bloch
sphere, a high-fidelity cloning is possible also for this regions on
the Bloch sphere.}

\section{\label{sec:3} Applying the meridional QCM for eavesdropping}

Although an eavesdropping of the (quantum) communication between two
partners, say Alice and Bob, might be seriously hampered by the
well-known impossibility to produce exact clones of quantum
information (non-cloning theorem), the generation of high-fidelity
copies may enlarge the success rate for such an attack. It is
therefore worth to know for both, the two partners who wish to
communicate as well as for a potential eavesdropper, how much our
suggested (meridional) QCM
(\ref{suggested-QCM-1})-(\ref{suggested-QCM-2}) will affect Quantum
Key Distribution (QKD) protocols. In this Section, we shall
therefore analyze with which success (or failure) an potential
eavesdropper, say Eve, may attack Bennett's B92 protocol
\cite{Bennett:92}.

In the B92 protocol, only two nonorthogonal quantum states are
utilized in order to encode and transmit the information about the
cryptographic key. As usual, we suppose that the information is sent
from Alice to Bob by means of a quantum communication channel.
\cor{At the beginning of the key distribution protocol, Alice
encodes each logical bit, 0 or 1, into two nonorthogonal states,
that can be parameterized in computational basis with single
parameter $\vartheta$ \cite{Ekert:94} as
\begin{eqnarray}
\label{B92_states}
   \ketm{u} & = & \cos{\frac{\vartheta}{2}} \ketm{0}
                + \sin{\frac{\vartheta}{2}}\ketm{1} \, ,
\nonumber\\[0.1cm]
   \ketm{v} & = & \sin{\frac{\vartheta}{2}} \ketm{0}
                + \cos{\frac{\vartheta}{2}}\ketm{1} \, .
\end{eqnarray}
The overlap of the states $O(\vartheta) \equiv |\sprm{u}{v}|^2 =
\sin^2{\vartheta}$ gives the distance between states $\ketm{u}$ and
$\ketm{v}$ in geometric sense. These qubits are then sent to Bob who
performs a positive operator-valued measurement (POVM), and the best
operators for that are \cite{Ekert:94}}
\begin{eqnarray}
\label{povm-1}
   G_1 &=& \frac{1}{1 \, + \, \sprm{u}{v}} \:
           \left( 1 - \ketm{u}\bram{u} \right) \, ,
   \\[0.1cm]
\label{povm-2}
   G_2 &=& \frac{1}{1 \, + \, \sprm{u}{v}} \:
           \left( 1- \ketm{v}\bram{v} \right) \, ,
   \\[0.1cm]
\label{povm-3}
   G_3 &=& 1 - G_1 - G_2 \, .
\end{eqnarray}
\cor{Only measurements with POVM elements $G_1$ and $G_2$ are
conclusive, because certain conclusion about received state
$\ketm{u}$ or $\ketm{v}$ can be made after the measurement.} After
all the qubits have been sent (and measured), Bob tells to Alice
numbers of conclusive measurements via a public channel, which can
be monitored but not modified by possible eavesdropper. Only those
bits (obtained in Bob's conclusive measurements) can be used to
construct the key, while all the rest need to be discarded because
no definite conclusion can be drawn from the outcome of Bob's
measurement. To test and recognize a (possible) eavesdropper, Alice
and Bob compare moreover the values of some of their bits via the
public channel in order to get an estimate how likely their
communication was disturbed.

In practice, a disturbance in the transmission of a (secret) key can
have very different reasons. Apart from an eavesdropper, the quantum
control during the preparation or transmission of the qubits might
be incomplete for a given realization of the quantum channel. For
all practical realizations of QKD protocols, therefore, a certain
error rate (disturbance) need to be accepted, and an eavesdropper
might be successful in extracting information even if the protocol
is inherently secure in the ideal case. To quantify the disturbance
in the transmission of a single qubit, a convenient measure is the
probability that Alice and Bob detect an error. If Bob would know
the state \ket{s} of one or several qubits in advance, that were
sent to him by Alice, he could easily test for a possible
eavesdropping attack. In this case, he will receive in general the
qubits no longer in a pure but a mixed state that has to be
described in terms of its density matrix $\rho$. The
\textit{discrepancy} that is detected by Bob is given by
\begin{eqnarray}
\label{discrepancy}
   D \,=\, 1 \, - \, \sprm{s}{\rho|s} \, .
\end{eqnarray}
Since Bob knows the maximal discrepancy $D_{max}$ for the given
channel (due to the incomplete quantum control of the given
transmission), he could recognize an eavesdropping attack for $D >
D_{max}$ and discard the key accordingly.

A central question for Eve is of how much information she can
extract from the transmission of the key if the disturbance due to
his attack should be $D < D_{max}$. From the initial agreement
between Alice and Bob about the basis states which are to be chosen
randomly, Eve might know that Alice prepares the qubits in one of
the two states (\ref{B92_states}) with probability $p_i = 1/2
\;(i=0,1)$. Before Eve has measured a given qubit, her (degree of)
ignorance is given by Shannon's entropy $H \,=\, -\sum p_i \log_2
p_i = -\log_2(1/2)$. After the measurement, she increased her
knowledge about the system by decreasing this entropy, a measure
that is called the mutual information that Eve has acquired due to
the measurement. Obviously, Eve will try to obtain as much
information as possible keeping the discrepancy $D < D_{max}$.

In order to discuss how much Eve will affect the transmission (and
thus increase her knowledge about the transmitted information), we
must specify the circumstances under which the eavesdropping attack
is made. \cor{Let us suppose that Eve performs \textit{incoherent}
(i.e. individual) attack on the communication channel with a QCM.
Here, we shall not yet specify the QCM explicitly in order to enable
us to compare different QCM's below. According to the incoherent
strategy, Eve will copy (attack) each qubit independently as they
are sent from Alice to Bob.} As output of her cloning
transformation, she then obtains two copies of one of the two
possible states \cor{$\rho_{\ketm{u}}$ and $\rho_{\ketm{v}}$ (which
just correspond to the two input states (\ref{B92_states}))} with a
fidelity as defined by the given QCM. While Eve transmits one of her
copies further to Bob, she could measure the second copy following
the same procedure as Bob.

To calculate the mutual information between Alice and Eve that is to
be extracted from the eavesdropping, we can follow the procedure as
described in Ref.~\cite{Fuchs:96} and \cite{Peres:02}. Using the
POVM elements (\ref{povm-1})-(\ref{povm-3}), the probability for Eve
to obtain the outcome $\mu$ is
\begin{equation}
\label{Probability}
   P_{\mu i} \, = \, {\rm Tr}(G_\mu \rho_i) \, ,
\end{equation}
and where the operators $\rho_i$ refer to the two possible states
$\{\rho_{\ketm{u}},\,\rho_{\ketm{v}}\}$ of her copy. After the
measurement, when she has obtained a particular outcome $\mu$, the
\textit{posterior} probability $Q_{i\mu}$ that $\rho_i$ was prepared
by Alice is
\begin{equation}
\label{post-probability}
   Q_{i\mu} \, = \, \frac{P_{\mu i}p_i}{q_\mu} \, ,
\end{equation}
where $q_\mu=\sum_j P_{\mu j}p_j$, and $p_j=1/2$ is the probability
for sending the states \ket{u} and \ket{v}  within the B92 protocol.
With these probabilities, the Shannon entropy (which was $H=-\log_2
(1/2)= 1$ initially), becomes
\begin{equation}
   H_\mu \, = \, -\sum_i Q_{i\mu} \log Q_{i\mu} \, ,
\label{shannon entrophy}
\end{equation}
once the result $\mu$ was obtained, and hence the mutual information
is
\begin{equation}
   I \, = \, H \, - \, \sum_\mu q_\mu H_\mu \, .
\label{mutual information}
\end{equation}

\cor{
\begin{figure}
\begin{center}
\includegraphics[scale=0.8]{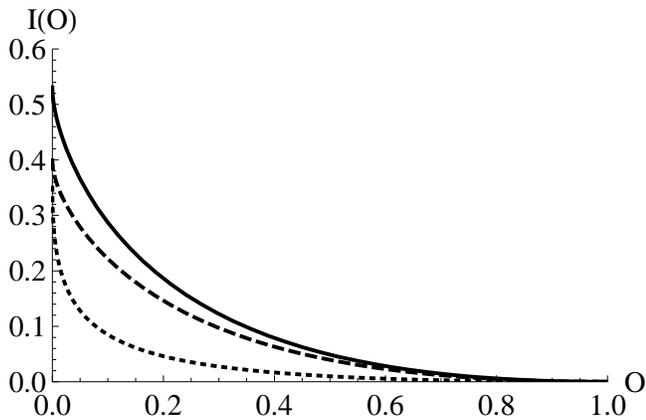}
\caption{The mutual information between Alice and Eve $I(O)$ as
function of the overlap $O$ of the states (\ref{B92_states}) for
eavesdropping with meridional (solid), equatorial (dashed) and
universal (dotted) QCM's.} \label{fig-2}
\end{center}
\end{figure}}

To determine the possible success of an eavesdropper, we only need
to analyze the explicit form of the output states $\rho_{\ketm{u}}$
and $\rho_{\ketm{v}}$ for a particular QCM. By substituting the
output states into Eqns.~(\ref{mutual information}) and
(\ref{discrepancy}) we may then calculate the mutual information and
discrepancy in case of an eavesdropping with the QCM. \cor{Since the
copies from the QCM are symmetric, the mutual information extracted
by Eve $I_{AE}$ equals to the mutual information obtained by Bob in
his measurements $I_{AB} = I_{AE} \equiv I$ and, obviously, depends
from the overlap of the states (\ref{B92_states}).}

\cor{If Eve applies \textit{universal} QCM, which provides copies
with fidelity $F=5/6$, she causes discrepancy $D\,=\,1/6\,\approx\,
0.17$ independently from the choice of the states
(\ref{B92_states}). The mutual information extracted by Eve is given
at Fig.~\ref{fig-2} with dotted line. For \textit{equatorial} QCM
with the fidelity $F = 1/2 + \sqrt{1/8}$ of the copies, discrepancy
equals $D = 1/2 - \sqrt{1/8}\,\approx\, 0.15$ for arbitrary states
(\ref{B92_states}) and the mutual information is shown at
Fig.~\ref{fig-2} with dashed line. For the eavesdropping with
\textit{meridional} QCM discrepancy depends from particular choice
of the states (\ref{B92_states}) and is given with inequality $0.05
\leq D \leq 0.10$. The mutual information in this case is shown at
Fig.~\ref{fig-2} with solid line. Apparently, our suggested QCM
introduces a lower disturbance into the data transmission between
Alice and Bob than it is caused by universal or equatorial QCM's. It
also enables Eve to extract in course of her eavesdropping more
information than obtained by means of these two QCM's.}

\cor{Suggested meridional QCM as well as universal and equatorial
QCM's belongs to the class of QCM's with an axillary system
\cite{Scarani:05}. There are, however, several proposals of optimal
cloning transformations without any axillary system
\cite{Scarani:05,Hillery:97,Bruss1:98}. It is, of course, important
to compare efficiencies of the eavesdropping with meridional QCM and
the best QCM without the axillary systems. Optimal cloning
transformation that clone at best two arbitrary pure states of a
qubit was developed in Refs.~\cite{Hillery:97,Bruss1:98}. This
transformation was shown to have extremely large fidelity $0.987
\leq F \leq 1$ between the input and output states. However, in this
type of cloning the copies are strongly entangled. That leads to the
fact that in the eavesdropping with the QCM without the axillary
system, Eve has very poor information about the result obtained by
Bob. Indeed, as the result of Bob's measurement on received qubit,
state of Eve's qubit reduces to strongly mixed state. In fact,
presence of the axillary system is necessary to eavesdrop
efficiently a QKD protocol based on single-qubit states
\cite{Scarani:05}.}

\cor{General question arises now, whether the incoherent
eavesdropping attack with meridional QCM discussed above is optimal
within B92 QKD protocol? To answer the question, we need to analyze
the success of the eavesdropper in line of alternative incoherent
attacks as well as alternative strategy for eavesdropping, namely,
\textit{coherent} ( or collective) attack when the eavesdropper has
opportunity to perform collective measurements on intercepted
qubits. Several incoherent attacks on the protocol have been
proposed before \cite{Ekert:94} such as \textit{intercept-resend}
attack and the \textit{attack with entangled probe}. It was found
that the efficiency of the eavesdropping depends from the overlap of
the states (\ref{B92_states}). Recently we showed that for wide
range of the overlap $ 0.07 \leq O(\vartheta) \leq 0.50$ of the
states \ket{u} and \ket{v}, the attack with meridional QCM gives
advantage for the eavesdropper to obtain more information causing
less discrepancy than any known incoherent attack \cite{Siomau:10}.
For small overlap when the states are almost orthogonal, however,
intercept-resend attack is optimal; while for large overlap the
attack with the entangled probe is optimal (see \cite{Siomau:10} for
further discussion). Moreover, the eavesdropping with meridional QCM
is optimal independently from incoherent or coherent strategy of the
eavesdropping, since it was shown that coherent strategy for the
eavesdropping gives negligible small additional information
comparing to the incoherent strategy in the protocols for quantum
key distribution based on single-qubit states
\cite{Scarani:05,Bechmann:99}.}

\section{\label{sec:4} Conclusions}

Unlike the well-known \textit{universal} \cite{Buzek:96} and
\textit{equatorial} \cite{Bruss:00} quantum cloning, we have
presented a QCM that provides high-fidelity copies for all states
from a selected meridian (i.e. half-circle) of the Bloch sphere.
\cor{This (so-called) meridional QCM was constructed to provide
hight-fidelity copies with $0.95 \,\geq\, F \,\geq\, 0.90$ for all
states along the Eastern meridian.} Although this QCM provides
high-fidelity copies for the Eastern meridian, it can be applied
with little adaptions also to other meridians. All what is needed to
follow the `optimization' procedures as described in
Subsection~\ref{subsec:2.c}.

The suggested QCM has been applied also to analyze a possible
eavesdropping attack in the data transmission between Alice and Bob,
following Bennett's B92 QKD protocol \cite{Bennett:92}. \cor{From
this analysis, it is shown that Eve, the eavesdropper, can obtain
more information from the meridional than from the universal or
equatorial QCM's. The probability that Bob (as the legitime user)
will detect the attack ($0.05 \leq D \leq 0.10$) is lower for the
meridional than in case of the universal ($D\approx0.17$) or
equatorial ($D\approx0.15$) quantum copying. Moreover, the
eavesdropping attack with meridional QCM is found to be optimal for
particular choices of states which are used to encode information in
the protocol \cite{Siomau:10}.}

\section*{Acknowledgments} M.S. thanks Sergei Kilin for discussions and
useful feedback. This work was supported by the DFG under the
project No. FR 1251/13.

\end{document}